\documentclass[11pt]{article}
\usepackage{arxiv}

\usepackage[T1]{fontenc}
\usepackage[utf8]{inputenc}
\usepackage{lmodern}
\usepackage{amsmath,amssymb,amsfonts}
\usepackage{graphicx}
\usepackage{textcomp}
\usepackage{xcolor}
\usepackage{booktabs}
\usepackage{bm}
\usepackage{verbatim}
\usepackage{url}
\usepackage{array}
\usepackage{multirow}
\usepackage{adjustbox}
\usepackage{float}
\usepackage[numbers,sort&compress]{natbib}
\usepackage[hidelinks]{hyperref}

\newcommand{\parencite}{\citep}
\newcommand{\textcite}{\citet}

\newcolumntype{L}[1]{>{\raggedright\arraybackslash}p{#1}}
\newcolumntype{C}[1]{>{\centering\arraybackslash}p{#1}}
\newcommand{\yes}{Yes}
\newcommand{\no}{No}
\newcommand{\act}[1]{\texttt{#1}}

\renewcommand{\shorttitle}{ZT-AIR}
\providecommand{\And}{\and}
\newenvironment{icistable}[3]{%
  \begin{table}[!htbp]\centering\caption{#2}\label{#1}%
  \begin{adjustbox}{max width=\textwidth}%
  \begin{tabular}{#3}\hline
}{%
  \hline\end{tabular}\end{adjustbox}\end{table}%
}
\newenvironment{icisfigure}[2]{%
  \begin{figure}[!htbp]\centering\caption{#2}\label{#1}%
}{\end{figure}}
\newcommand{\safeincludegraphics}[2][]{%
  \IfFileExists{#2}{\includegraphics[#1]{#2}}{%
    \fbox{\parbox{0.86\linewidth}{\centering Missing image: \texttt{\detokenize{#2}}}}%
    \typeout{WARNING: Missing figure file #2}%
  }%
}
\date{}
\begin{document}

\title{Zero Trust-Enabled Adaptive Incident Response using Hierarchical Reinforcement Learning}

\author{
Jisoo Lee \\
\textit{Indiana University} \\
Bloomington, IN, USA \\
\And 
Bhagath Singh Bondili \\
\textit{Indiana University} \\
Bloomington, IN, USA \\
\And 
Harshith Doppalapudi \\
\textit{Indiana University} \\
Bloomington, IN, USA \\
\And 
Gwynith Godin \\
\textit{United States Military Academy} \\
West Point, NY, USA \\
\And 
Nathaniel D. Bastian \\
\textit{United States Military Academy} \\
West Point, NY, USA \\
\And 
Ankit Shah\thanks{Send correspondence to ankit@iu.edu} \\
\textit{Indiana University} \\
Bloomington, IN, USA \\
}
\maketitle

\begin{abstract}
Modern enterprises increasingly operate through complex information systems (IS) enabled by cloud computing, remote and hybrid work practices, and digitally integrated supply chains. These structural shifts have expanded organizational attack surfaces and weakened traditional perimeter-based trust assumptions, giving rise to new and persistent forms of cyber threats. In response, adversaries have shifted toward identity-centric, trust-based intrusion techniques that exploit legitimate access within organizational systems. This evolution has created a pronounced speed asymmetry, whereby attackers can adapt and act at machine speed while defenders remain constrained by reactive or rule-based incident response (IR) processes. Drawing on design science research, we propose ZT-AIR, an adaptive IR artifact that uses zero trust architecture as its design theory and hierarchical reinforcement learning as its kernel theory. ZT-AIR learns context-aware trust decisions, separates strategic prioritization from tactical execution, encodes zero-trust controls as actions, and jointly optimizes security effectiveness and operational continuity. Experimental results across diverse adversarial strategies, network scales, and defense environments show that ZT-AIR achieves a strong balance between security and continuity through frequent, minimally disruptive interventions. Practitioner validation further confirms the design’s practical viability. This study contributes to IS research by extending computational design science from threat identification to response execution, theorizing zero trust architecture for adaptive IR, and deriving reusable design principles for adaptive decision support artifacts.
\end{abstract}


Contemporary organizations operate in highly complex sociotechnical environments, where cloud services, hybrid workforces, and third-party integrations create dense hyper-connections. This complexity produces wicked problems that resist traditional solutions \parencite{benbya2020complexity}. Cybersecurity has emerged as one such problem because digital interconnections expand the organizational attack surface and blur previously well-defined trust boundaries. Attacks are no longer dominated by overt ``break-ins.'' Instead, adversaries increasingly exploit compromised credentials, lateral movement, and trust-based mechanisms to conduct stealthy intrusions, effectively ``logging in'' to enterprise systems \parencite{crowdstrike2025global}. Trust-based attacks constitute nearly 25\% of all initial access vectors leading to data breaches \parencite{ibm2025cost}.

A critical speed asymmetry further exacerbates the defender's challenge. Adversaries initiate lateral movement within an average of 48 minutes and achieve data exfiltration within two days \parencite{crowdstrike2025global,paloalto2025ciso}. Defenders, in contrast, require more than 240 days on average to contain trust-based breaches \parencite{ibm2025cost}. This disparity renders static, rule-based incident response (IR) insufficient. It necessitates a shift toward adaptive IR, where defensive actions can be executed in near real time, adapt to unfolding attack behavior, and improve through experience.

Two bodies of work offer partial but incomplete answers to this challenge. The first is zero trust architecture (ZTA), which has emerged as the authoritative organizational response to trust boundary erosion \parencite{nist2025zta,nist2020zta}. ZTA assumes breach as the default threat posture and enforces continuous, context-aware verification across all users, devices, and workloads. Crucially, its architectural design elements provide a principled vocabulary for adaptive IR \parencite{nist2025zta,nist2020zta}: (1) foundational tenets that assume breach and require continuous verification of every access request, (2) an adaptive trust evaluation logic that weighs identity, device posture, and behavioral signals against security and operational continuity objectives, (3) a hierarchical decision structure that separates strategic trust judgments from the tactical directives that carry them out and from their enforcement at the resource boundary, and (4) fine-grained enforcement mechanisms such as access verification, access revocation, and micro-segmentation. Yet ZTA adoption remains limited, as echoed in our consultation with security operations center (SOC) practitioners. The barrier is the absence of an approachable computational model.

The second body of work is reinforcement learning (RL)-based adaptive IR, in which agents learn defensive policies through sequential interaction with adversarial environments \parencite{hurten2024hierarchical,kiely2309autonomous,mukherjee2025large,ren2025arcs,singh2024hierarchical,wei2024offline}. Yet existing RL-based IR models typically treat ZTA as an implicit environmental assumption rather than an explicit design theory shaping the agent's decision structure.

Taken together, these two bodies of work leave a consequential design question unaddressed: \textit{How can ZTA be operationalized as a structured, learnable decision process to enable adaptive IR in dynamic adversarial environments?} This question also surfaces a gap in IS Computational Design Science Research (DSR). Existing IS cybersecurity studies have largely focused on AI-enabled threat intelligence, particularly for threat identification and classification (see Table~\ref{tab:isdsr}). These contributions support the preparatory phases of the National Institute of Standards and Technology (NIST) IR lifecycle \parencite{nist2025ir} but leave the response phase unaddressed. This imbalance constrains continuous improvement, which requires learning across all IR phases \parencite{rezazade2022organizations}. To our knowledge, no IS DSR exists that translates ZTA's normative principles into an executable, learning-based IR artifact.

This study addresses these gaps by adopting ZTA as a design theory and hierarchical reinforcement learning (HRL) as its kernel theory. HRL decomposes complex decision problems into a high-level policy that selects strategic objectives over longer time horizons and a low-level policy that executes concrete actions to fulfill those objectives \parencite{dayan1992feudal,dietterich2000hierarchical,sutton1999between}. This shared emphasis on hierarchical, adaptive, sequential decision-making grounds a principled mapping between ZTA and HRL. On this foundation, we propose Zero Trust-enabled Adaptive IR (ZT-AIR), a design science artifact that operationalizes ZTA as an adaptive IR system.

We evaluate ZT-AIR against a progressive set of baselines, namely a rule-based agent, flat deep RL (DRL) without hierarchy, and HRL without ZT actions, across two complementary testbeds. Cyberwheel \parencite{oesch2024cyberwheel} is a centralized single-agent environment. The Cyber Autonomy Gym for Experimentation challenge 4 (CAGE4) environment \parencite{kiely2025cage4} is a distributed multi-agent environment. The results reveal that operationalizing ZTA as an executable decision structure produces a joint advantage in security effectiveness and operational continuity. This advantage remains robust to unseen adversary strategies, network scales, and defense operation settings.

This study makes three contributions to IS research. First, it extends IS computational design science to the response execution phase by proposing the first IS computational artifact for adaptive IR. Second, it establishes ZTA as an IS design theory by translating ZTA's hierarchical structure, trust evaluation, and enforcement mechanisms into the computational elements of an HRL agent. Third, controlled ablation across progressively more capable baselines demonstrates that security effectiveness and operational continuity are jointly achievable, not competing, objectives when ZTA principles are faithfully operationalized.

\section{Literature Review}

\subsection{IS Computational Design Science Research on Cyber Defense}

IS computational DSR, which emphasizes the purposeful creation and rigorous evaluation of IT artifacts \parencite{rai2017editor}, has made significant strides in the cyber defense area. As summarized in Table~\ref{tab:isdsr}, studies published in premier IS journals leverage AI and analytics to support threat intelligence, vulnerability assessment, defensive robustness, and fraud detection. Yet the distribution of this work across the IR life cycle \parencite{nist2025ir} reveals a pronounced and persistent imbalance. Of the 21 studies reviewed, 19 address Identify, Protect, and Govern functions in the preparatory phase, while no study addresses the Respond function in the execution phase. This pattern indicates a structural rather than transitional gap.

\begingroup\footnotesize
\begin{icistable}{tab:isdsr}{IS Computational Design Science Literature on Cyber Defense by IR Function}{|L{0.7in}|L{1.75in}|C{0.25in}|L{3.05in}|}
\textbf{IR function} & \textbf{Definition} & \textbf{n} & \textbf{Studies} \\ \hline
\multicolumn{4}{|l|}{\textit{Preparation phase}} \\ \hline
Govern & Establish, communicate, and monitor cybersecurity risk management strategy, expectations, and policy & 1 & \textcite{benjamin2019dice} \\ \hline
Identify & Understand the organization's current cybersecurity risks & 15 & \textcite{ampel2025computational,ampel2024creating,ampel2024improving,benjamin2016examining,biswas2022text,ebrahimi2020semi,ebrahimi2022cross,jain2025vulnerability,kumar2025deepsecure,li2022discovering,li2016identifying,samtani2017exploring,samtani2022linking,sun2019regulating,ullman2024enhancing} \\ \hline
Protect & Use safeguards to manage cybersecurity risks & 3 & \textcite{ebrahimi2025radar,ebrahimi2025defending,sharma2020paradox} \\ \hline
\multicolumn{4}{|l|}{\textit{Execution phase}} \\ \hline
Detect & Find and analyze possible attacks and compromises & 2 & \textcite{ampel2026automatically,du2025follow} \\ \hline
Respond & Take actions regarding a detected incident & 0 & --- \\ \hline
Recover & Restore assets and operations affected by an incident & 0 & --- \\
\end{icistable}
\endgroup

The root cause of this imbalance is the absence of a design theory grounding computational artifacts for adaptive IR. IR is a sequential decision-making problem under adversarial uncertainty. Such a problem requires a cybersecurity design theory that specifies how trust is dynamically evaluated, what response actions are permissible, and how security objectives are balanced against operational continuity. Prior IS research has not supplied this design foundation.

\subsection{Cybersecurity Incident Response Studies Using Reinforcement Learning}

The broader cybersecurity IR literature has modeled IR as a sequential decision-making problem and addressed it using RL. RL agents learn optimal policies through continuous interaction with an environment, accounting for how current actions influence future states \parencite{sutton1998reinforcement}. Recent studies (Table~\ref{tab:rlir}) are evaluated in simulated environments, principally the Cyberwheel framework and the CAGE testbeds. As simulated environments have grown in scale and realism, flat DRL has proven insufficient due to the curse of dimensionality. This limitation has motivated a shift toward HRL, which decomposes large state and action spaces into tractable subproblems.

\begingroup\footnotesize\setlength{\tabcolsep}{4pt}
\begin{icistable}{tab:rlir}{Recent Incident Response Studies Using Reinforcement Learning}{|L{1.02in}|L{0.75in}|C{0.55in}|L{1.1in}|C{0.35in}|L{1.08in}|C{0.67in}|}
\textbf{Study} & \textbf{Testbed} & \textbf{Trust-based attack} & \textbf{Defense actions} & \textbf{RL type} & \textbf{Algorithm} & \textbf{Bi-objective reward} \\ \hline
\textcite{mukherjee2025large} & Cyberwheel & \yes & Conventional only & DRL & PPO & \no \\ \hline
\textcite{ren2025arcs} & Incident dataset & \yes & Conventional only & HRL & Dual-stream deep Q-network & \yes \\ \hline
\textcite{singh2024hierarchical} & CAGE4 & \yes & Conventional only & HRL & Multi-agent PPO & \yes \\ \hline
\textcite{hurten2024hierarchical} & CAGE4 & \yes & Conventional only & HRL & Multi-agent PPO & \yes \\ \hline
\textcite{wei2024offline} & CyberVAN & \yes & Conventional only & DRL & Model-based value iteration & \yes \\ \hline
\textcite{kiely2309autonomous} & CAGE2 & \yes & Conventional only & HRL & PPO & \no \\
\end{icistable}
\endgroup

Despite these technical advances, a critical gap persists in the treatment of ZTA. Most studies adopt threat scenarios where adversaries have already gained initial access and penalize defensive costs or benign-user disruption in the reward. Yet none operationalizes ZTA's core control elements. Key mechanisms (e.g., access verification, access revocation, and micro-segmentation) are absent from the defensive action spaces.

\section{Research Gaps and Questions}

The preceding review reveals a dual gap. IS computational DSR has generated substantial knowledge on the preparatory functions of the IR life cycle, such as threat identification, but has not addressed the Respond function. Concurrently, RL-based IR research has advanced hierarchical agent design, yet treats ZTA as a contextual threat assumption rather than an explicit architectural design logic. To address these gaps, we design the ZT-AIR artifact, which translates ZTA principles into an adaptive, learnable IR decision structure, and evaluate it through the following research questions. \textbf{RQ1.} What are the performance contributions of the operationalized ZTA design elements in ZT-AIR to security effectiveness and operational continuity? \textbf{RQ2.} To what extent does ZT-AIR maintain its effectiveness against unseen adversarial strategies outside the training distribution? \textbf{RQ3.} To what extent does ZT-AIR maintain its effectiveness as network scale increases? \textbf{RQ4.} Does ZT-AIR remain effective when extended to a distributed multi-agent defense setting?

\section{Theoretical Foundations}

ZTA serves as a \textit{design theory} that specifies what adaptive IR should achieve. HRL serves as a \textit{kernel theory} that specifies how that logic can be learned and executed under the sequential, uncertain, and partially observable conditions of real-world IR \parencite{gregor2013positioning}.

\subsection{Zero Trust Architecture as Design Theory}

ZTA rejects implicit trust and mandates continuous, context-aware verification of every access request \parencite{nist2025zta,nist2020zta}. As a design theory, ZTA supplies four elements that directly govern artifact design. The first is its \textit{foundational tenets}, which assume breach as the baseline threat posture and treat continuous verification as a non-negotiable principle. The second is its \textit{adaptive trust evaluation logic}. Every access decision is mediated by a trust algorithm that assesses trustworthiness using identity, device posture, and behavioral signals, and that must balance security effectiveness against operational continuity. The third is its \textit{hierarchical decision structure}. The Policy Engine (PE) renders strategic trust judgments, the Policy Administrator (PA) translates them into enforcement directives, the Policy Decision Point (PDP) coordinates PE--PA interaction, and the Policy Enforcement Point (PEP) executes those directives at the resource boundary. The fourth is its \textit{enforcement mechanisms}. Enhanced identity governance instantiates credential access control through access verification and access revocation, while micro-segmentation enforces granular network boundaries to contain unauthorized traffic.

Critically, ZTA prescribes that the trust algorithm must be dynamic and context-sensitive, while leaving its computational logic unspecified \parencite{nist2025zta,nist2020zta}. Operationalizing ZTA for adaptive IR therefore requires a computational mechanism that learns and continuously refines trust evaluation policies from environmental feedback, rather than predefined rules.

\subsection{Hierarchical Reinforcement Learning as Kernel Theory}

HRL addresses this operationalization requirement by decomposing complex sequential decision problems into multiple levels of abstraction \parencite{dayan1992feudal,dietterich2000hierarchical,sutton1999between}. This decomposition directly instantiates ZTA's hierarchical decision structure. The high-level policy corresponds to the PE's strategic trust judgments, and the low-level policy corresponds to the PA's tactical command directives. Their subgoal communication corresponds to the PDP's PE--PA coordination, and the environment actuation interface corresponds to the PEP. The trust algorithm itself is instantiated as a learned decision logic approximated via proximal policy optimization \parencite[PPO;][]{schulman2017proximal}, rather than as manually designed trust rules. Table~\ref{tab:mapping} formalizes this ZTA--HRL mapping and derives the artifact instantiation requirements that govern ZT-AIR's design.

\begingroup\footnotesize\setlength{\tabcolsep}{4pt}
\begin{icistable}{tab:mapping}{ZTA-to-HRL Translation and Artifact Instantiation Requirements}{|L{2.0in}|L{1.7in}|L{2.4in}|}
\textbf{ZTA concept} & \textbf{HRL interpretation} & \textbf{Artifact instantiation requirements} \\ \hline
\textbf{A. Foundational tenets:} assume breach (no implicit trust); continuous verification (every session is re-evaluated) & Persistent, adversarial, non-stationary environment; sequential policy evaluation over time & Train under ongoing intrusion attempts with sustained lateral movement pressure; recurrent policy updates at each time step \\ \hline
\textbf{B. Adaptive trust learning and evaluation:} trust algorithm (dynamic context-aware decision logic); security--operational continuity balance & Learned sequential decision logic that continuously evaluates evolving context; bi-objective reward signal & Learn and recurrently update policies that approximate risk-adaptive trust decisions under uncertainty; jointly optimize attack suppression and operational continuity \\ \hline
\textbf{C. Hierarchical decision structure:} Policy Engine (strategic trust decisions); Policy Administrator (tactical command directives); Policy Decision Point (PE--PA coordination); Policy Enforcement Point (action implementation) & High-level policy (long-horizon planning); low-level policy (short-horizon planning); hierarchical policy coordination; action implementation interface & Select strategic IR objectives (e.g., where to prioritize defensive attention); translate objectives into concrete IR actions (e.g., what to execute); coordinate PE--PA interaction through goal/option communication; execute actions via the testbed actuation layer \\ \hline
\textbf{D. Enforcement mechanisms:} enhanced identity governance (continuous authentication and privilege management); micro-segmentation (granular network controls to limit lateral movement) & Identity-level access control actions; network-level isolation actions with graduated containment intensity & Implement access verification and access revocation with minimal cost structure; implement micro-segmentation with lower disruption cost than full isolation \\
\end{icistable}
\endgroup

\section{Artifact Design: ZT-AIR}

We first formalize IR as a Partially Observable Markov Decision Process (POMDP), and then instantiate the artifact by translating ZTA concepts into HRL's learning components. Figure~\ref{fig:framework} provides a visual overview of the proposed artifact.

\begin{icisfigure}{fig:framework}{The ZT-AIR Artifact}
\safeincludegraphics[width=6.2in]{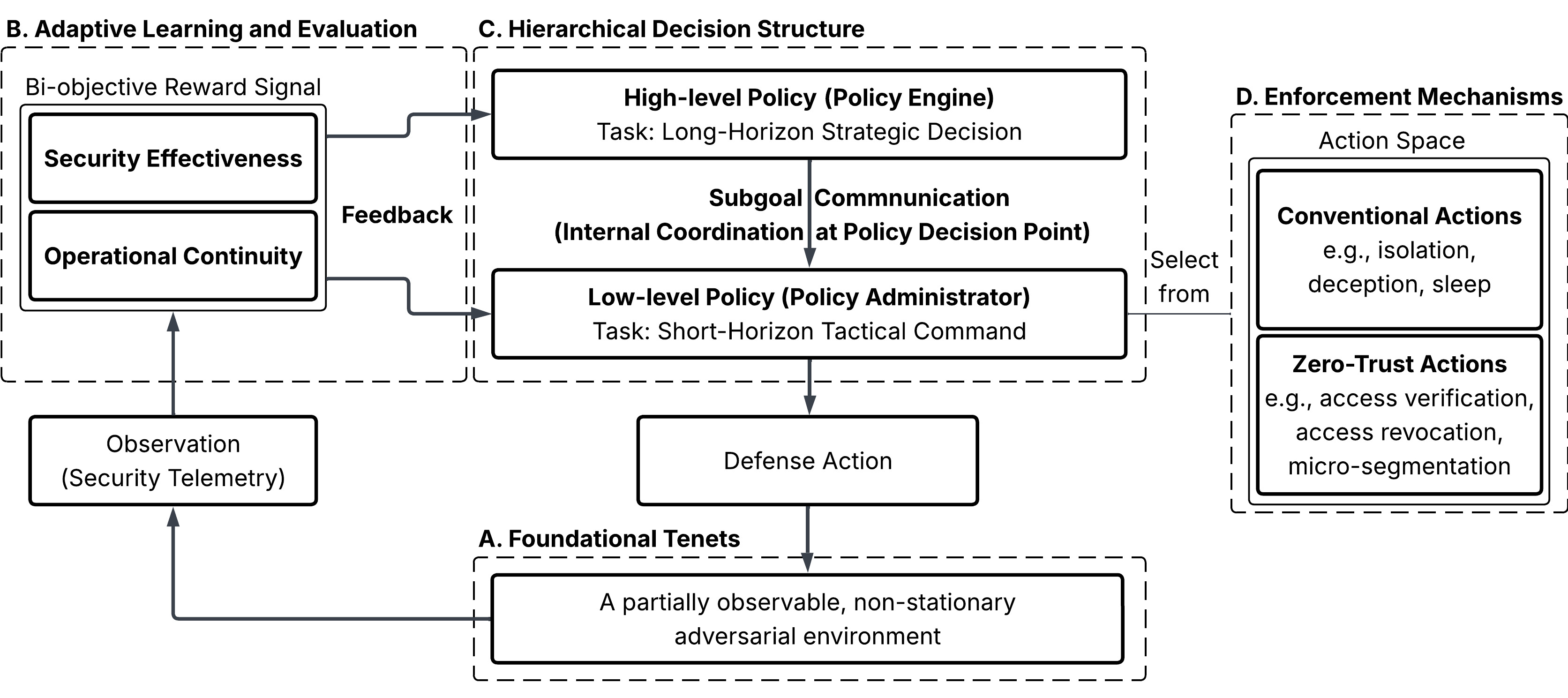}
\end{icisfigure}

\subsection{Adaptive Incident Response as a Partially Observable Markov Decision Process}

In real-world security operations, defenders cannot directly observe the network state. They must instead infer attack progression from noisy, incomplete signals (e.g., alerts, logs, and local telemetry). Adaptive IR is thus a sequential decision-making problem under uncertainty. We formalize IR as a POMDP $\mathcal{M}=(\mathcal{S},\mathcal{O},\mathcal{A},P,R,\gamma)$. The defender selects actions $a_t\in\mathcal{A}$ based on observations $o_t\in\mathcal{O}$ derived from security telemetry rather than the hidden state $s_t\in\mathcal{S}$, and optimizes expected long-term returns under reward $R$ and discount factor $\gamma$. The learned object is the policy, $\pi$, that determines how defensive controls should be selected, sequenced, and adapted as threats evolve. ZT-AIR learns this policy via PPO-based HRL.

\subsection{ZT-AIR Design Instantiation}

ZT-AIR instantiates each POMDP element to fulfill a ZTA-to-HRL requirement from Table~\ref{tab:mapping}, as indicated by the dimension labels below.

\textbf{State and Observation Space.} The underlying environment state $s_t$, the complete configuration of the enterprise and adversary, is not directly observable. The agent therefore operates on a belief state $b_t(s)=P(s_t=s\mid o_{1:t},a_{1:t-1})$ and acts on observations constructed from two information streams, $o_t=[x^{E}_t;\,x^{A}_t]$, where $x^{E}_t$ denotes the observable enterprise context and $x^{A}_t$ captures observed or inferred adversarial activity. This dual-stream structure operationalizes ZTA's foundational tenets of assume breach and continuous verification and its trust algorithm (Dimensions A and B).

\textbf{Action Space.} The action space $\mathcal{A}$ integrates two categories of defensive controls. ZTA actions derived from enforcement mechanisms (Dimension D) include, for example, access verification, access revocation, and micro-segmentation, while conventional IR actions include deception and isolation.

\textbf{Reward Function.} The reward function encodes ZTA's security effectiveness--operational continuity balance (Dimension B):
\begin{equation}
R_t=R_{\mathrm{security},t}+R_{\mathrm{operation},t}.
\label{eq:reward}
\end{equation}
$R_{\mathrm{security},t}$ penalizes attack escalation (e.g., lateral movement, privilege escalation, and final impact) and rewards containment and mitigation. $R_{\mathrm{operation},t}$ penalizes unnecessary costs from defensive actions (e.g., excessive isolation, unwarranted access revocation, and service interruption).

\textbf{Hierarchical Policy.} ZT-AIR trains two PPO policies that mirror ZTA's hierarchical decision structure (Dimension C). The high-level policy $\pi_H$ plays a PE role, selecting an abstract defense objective $z_t$ over a long temporal horizon. The low-level policy $\pi_L$ plays a PA role, selecting an executable action $a_t$ conditioned on the current belief state $b_t$ and the selected option $z_t$:
\begin{equation}
z_t\sim\pi_H(z_t\mid b_t),\qquad a_t\sim\pi_L(a_t\mid b_t,z_t).
\end{equation}
PDP coordination is realized through subgoal communication from $\pi_H$ to $\pi_L$. This hierarchical decision structure also operationalizes ZTA's trust algorithm (Dimension B).

\section{Experimental Design and Evaluation}

We evaluate ZT-AIR through three primary experiments in Cyberwheel, a centralized single-agent testbed, addressing RQ1--3. An extended experiment in CAGE4 assesses whether its performance advantage generalizes to distributed multi-agent defense conditions (RQ4). Our design choices were consulted with SOC practitioners, including a manager, a senior analyst, and two senior engineers.

\subsection{Common Evaluation Setup}

\textbf{Baselines.} We compare ZT-AIR against three baseline models forming a progressive ladder that isolates the contribution of each ZTA design element. (1) A rule-based approach, validated by a collaborating SOC, selects defense actions reactively and without learning. Its predefined rule deploys decoys, isolates high-alert hosts, and otherwise does nothing. (2) The DRL agent learns a flat PPO-based policy that maps observations directly to conventional defense actions without hierarchical decomposition. (3) The HRL agent introduces hierarchical decomposition (same as ZT-AIR) but remains restricted to conventional defense actions, without ZTA enforcement actions.

\textbf{Evaluation Metrics.} Five metrics jointly capture security effectiveness and operational continuity. (1) \textit{Compromised rate} measures the averaged percentage of successful lateral movement actions on real (non-decoy) hosts relative to total attack attempts. (2) \textit{Impact rate} measures the averaged percentage of successful impact actions, which occur only after lateral movement succeeds, on real hosts relative to total attack attempts. (3) \textit{Defensive success rate} measures the defender's ability to neutralize adversarial actions. It is defined in Cyberwheel as the percentage of adversarial actions either blocked directly or diverted to decoys and in CAGE4 as the proportion of successful impact out of the lateral movement. (4) \textit{Mean time to respond} (MTTR) is total time steps divided by the number of defensive interventions, where smaller values indicate more frequent, responsive defense. (5) \textit{Operational impact} captures the average of defense action cost (Cyberwheel) or disrupted benign activity (CAGE4), where lower values indicate better operational continuity.

\textbf{Action Analysis.} We also report defense action distributions (the proportion of each action type used across episodes) to link learned behavior to the intended trade-off.

\subsection{Cyberwheel: Primary Analytical Testbed}

\textbf{Environment Settings and ZT-AIR Implementation.} Cyberwheel \parencite{oesch2024cyberwheel} is a simulation testbed developed by the U.S. Department of Energy's Oak Ridge National Laboratory. The attacker enters via compromised credentials in a user subnet and progresses through a multi-stage kill chain of reconnaissance (ping sweep and port scan), lateral movement, discovery, privilege escalation, and impact \parencite{MITRE2024}. Defense is centralized and single-agent, with the defender inferring attack progression from detector-generated alerts and minimal context. ZT-AIR is accordingly implemented as a centralized hierarchical defender. The high-level policy $\pi_H$ selects a subnet to target, while the low-level policy $\pi_L$ selects tactical enforcement actions within the chosen focus region.

The action space supports both conventional controls and ZTA actions, as defined in Table~\ref{tab:cwactions}. Whereas conventional isolation severs legitimate communication, ZTA-derived actions enable selective containment (e.g., reducing trust and restricting lateral connectivity while preserving authorized communication). SOC practitioners reviewed the action set and cost levels as a reasonable representation of relative disruption.

\begingroup\footnotesize
\begin{icistable}{tab:cwactions}{Defense Actions in Cyberwheel}{|L{0.85in}|L{0.9in}|L{3.2in}|C{0.85in}|}
\textbf{Type} & \textbf{Action} & \textbf{Description} & \textbf{Operational cost} \\ \hline
\multirow{3}{=}{Conventional} & Do nothing & Observe without modifying the network state. & None \\ \cline{2-4}
 & Deploy decoy & Deploy a deceptive host to attract attacker and generate alerts. & Medium \\ \cline{2-4}
 & Isolate host & Fully disconnect a host from the network. & High \\ \hline
\multirow{3}{=}{Zero Trust} & Verify access & Re-authenticate access to block credential-dependent movement (e.g., lateral movement and discovery). & Medium \\ \cline{2-4}
 & Revoke access & Remove outbound access privileges to constrain post-compromise activity (e.g., discovery, privilege escalation, and impact). & Medium \\ \cline{2-4}
 & Micro-segment & Enforce inbound access controls to restrict unauthorized credential-dependent actions while preserving approved communication. & Low \\
\end{icistable}
\endgroup

The reward implements Equation~\ref{eq:reward} in Cyberwheel as follows. $R_{\mathrm{security},t}$ comprises an attack-outcome term and a defense-outcome term. The attack-outcome term penalizes detected kill-chain escalation weighted by zone criticality (user $<$ demilitarized zone (DMZ) $<$ server), and the defense-outcome term reflects immediate defensive effectiveness. $R_{\mathrm{operation},t}$ is a defense-cost term that proxies the disruption of operational continuity through predefined action and recurring control costs, because the operational availability (e.g., benign user traffic) in Cyberwheel is not directly observable.

\textbf{Experimental Design.} The environment provides two categories of test conditions. Three adversary strategies vary in propagation behavior and attack goals. Random search (RS, the default) selects target hosts uniformly at random. Breadth-first search (BFS) expands laterally across adjacent hosts before progressing deeper. Depth-first search (DFS) pursues a single targeted path toward high-value assets. Three network scales expand the state and action spaces, namely small (the default, with user, DMZ, and server subnets and 15 hosts), medium (7 subnets and 100 hosts), and large (20 subnets and 1,010 hosts). Each evaluation comprises 50 episodes of 100 time steps. \textit{RQ1 -- ZTA Operationalization Value.} All four models are trained and tested on the small-network RS setting, and statistical significance is assessed via paired $t$-tests between adjacent models across all five metrics. \textit{RQ2 -- Threat Strategy Robustness.} All four models trained under RQ1 conditions are tested against unseen strategies (BFS, DFS) without retraining. \textit{RQ3 -- Enterprise-Scale Applicability.} All four models are independently trained and tested across three network scales, and cross-scale performance degradation is compared.

\subsection{CAGE4: Extended Analytical Testbed}

\textbf{Environment Settings and ZT-AIR Implementation.} CAGE4 \parencite{kiely2025cage4} is a cyber defense testbed co-sponsored by the Defense Advanced Research Projects Agency (DARPA) and Department of War (DoW). The attacker enters via a supply-chain or contractor pathway outside initial defender-team coverage and progresses through a high-level kill-chain structure. Three conditions shape the implementation of ZT-AIR. First, each defense agent in CAGE4 is assigned to an individual defensive region. ZT-AIR therefore implements $\pi_H$ as meta-strategy selection, the choice of a high-level defensive intent (Investigate, Traffic Control, and Recover), and $\pi_L$ as tactical action selection within each agent's region. Second, ZT-AIR extends the native CAGE4 action space by adding three ZTA enforcement controls to the environment's existing conventional actions. Each ZTA action is mapped into a meta-strategy in a manner consistent with its defensive purpose (Table~\ref{tab:cage4actions}). Third, CAGE4 provides local, traffic-derived observations within each agent's assigned zone, and operational continuity is directly observable through the success and failure of benign user-traffic activities. The reward preserves the same bi-objective structure. $R_{\mathrm{security},t}$ penalizes successful attacks on defended hosts and rewards attack removal, with severity adjusted according to the scenario stage of CAGE4. $R_{\mathrm{operation},t}$ penalizes disruption of benign user activities (e.g., access-service failures) and discourages unnecessary defensive actions on uncompromised hosts.

\begingroup\footnotesize
\begin{icistable}{tab:cage4actions}{Defense Actions in CAGE4}{|L{0.8in}|L{0.85in}|L{1.05in}|L{3.1in}|}
\textbf{Strategy} & \textbf{Type} & \textbf{Action} & \textbf{Description} \\ \hline
\multirow{4}{=}{Investigate} & \multirow{3}{=}{Conventional} & Monitor & Collect information on malicious activity in the subnet. \\ \cline{3-4}
 & & Analyze & Inspect a host to detect the attacker's presence. \\ \cline{3-4}
 & & Deploy decoy & Plant a honeypot on a host to attract an attacker. \\ \cline{2-4}
 & Zero Trust & Verify access & Re-authenticate a host to block credential-based lateral movement. \\ \hline
\multirow{3}{=}{Traffic Control} & \multirow{2}{=}{Conventional} & Block traffic zone & Block inbound contractor-to-subnet traffic. \\ \cline{3-4}
 & & Allow traffic zone & Re-permit previously blocked contractor-to-subnet traffic. \\ \cline{2-4}
 & Zero Trust & Micro-segment & Remove user-level sessions while preserving approved traffic. \\ \hline
\multirow{3}{=}{Recover} & \multirow{2}{=}{Conventional} & Remove & Terminate user-level sessions on a target host. \\ \cline{3-4}
 & & Restore & Rebuild a host to a known-good state, removing all compromises. \\ \cline{2-4}
 & Zero Trust & Revoke access & Immediately purge all sessions by invalidating credentials. \\
\end{icistable}
\endgroup

\textbf{Experimental Design.} \textit{RQ4 -- Operational Generalizability.} DRL, HRL, and ZT-AIR are trained and evaluated in CAGE4 on 50 episodes, each consisting of 500 time steps.

\section{Results}

\subsection{Experiment Set 1: ZTA Operationalization Value (RQ1)}

Figure~\ref{fig:exp1} summarizes model performance in the small-network RS setting. ZT-AIR delivers the strongest aggregate performance. It achieves the lowest compromised rate, impact rate, and operational impact and the highest defensive success rate while maintaining a low MTTR.

\begin{icisfigure}{fig:exp1}{Performance (Top) and Action Distributions (Bottom) in the Baseline Setting}
\safeincludegraphics[width=6.2in]{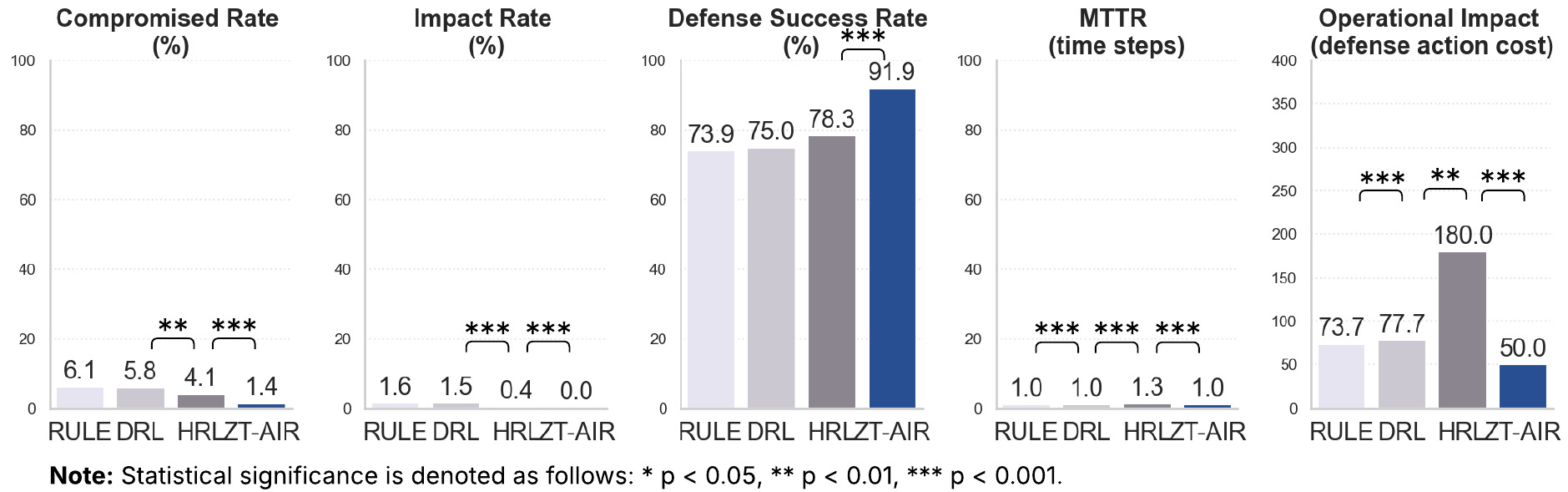}\\[4pt]
\safeincludegraphics[width=2.2in]{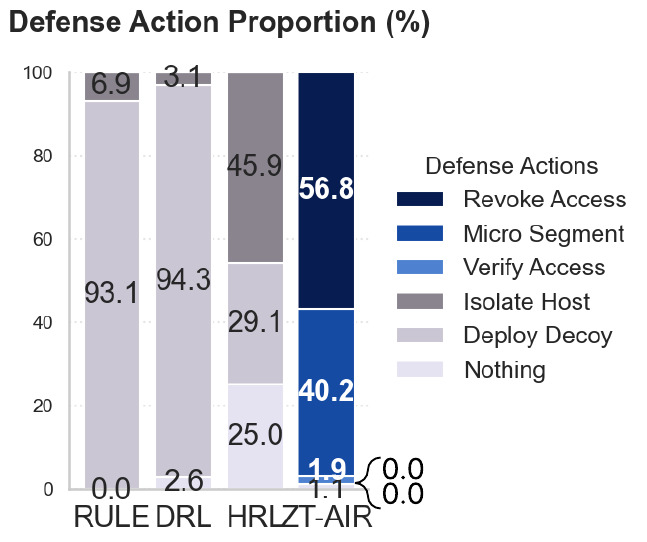}
\end{icisfigure}

\textbf{Adaptive learning (rule-based vs.\ DRL).} Adaptive learning alone yields limited improvement in this setting. DRL slightly improves compromised rate, impact rate, and defensive success rate, but none of these differences is statistically significant. The action distributions in Figure~\ref{fig:exp1} explain why these gains remain marginal. Both models rely heavily on deception, which a single flat policy managing all subnets and hosts finds locally optimal. The only significant differences appear in MTTR and operational cost. DRL learns to wait when no beneficial action is apparent, whereas the rule-based policy intervenes almost unconditionally.

\textbf{Hierarchical decision structure (DRL vs.\ HRL).} Hierarchical decomposition substantially improves security effectiveness through strategic targeting. Despite less frequent intervention (longer MTTR), HRL significantly reduces compromised rate and impact rate. The action distributions show that HRL shifts away from deception-centered defense and toward a more aggressive containment strategy centered on host isolation. However, because isolation blocks both malicious and legitimate traffic, this stronger containment comes at a substantial continuity cost. HRL more than doubles operational impact relative to DRL. Hierarchical structure thus enables more precise strategic execution, but when constrained to coarse conventional controls, it cannot escape the security--operation trade-off.

\textbf{ZTA enforcement mechanisms (HRL vs.\ ZT-AIR).} Enforcement mechanisms are the critical design element that transforms hierarchical decision into a precise and operationally sustainable defense strategy. ZT-AIR significantly improves all four effectiveness metrics, namely compromised rate, impact rate, defensive success rate, and MTTR, while reducing operational impact to only 28\% of HRL's level. Unlike HRL, ZT-AIR uses neither decoys nor full host isolation. It instead relies primarily on \act{revoke\_access} and \act{micro\_segment}, which enable frequent intervention across the kill chain while avoiding the operational burden of isolation. 
In sum, the advantage of ZT-AIR is not attributable to adaptive learning alone, but to the operationalization of hierarchical decision structure and enforcement mechanisms in ZTA.

\subsection{Experiment Set 2: Threat Strategy Robustness (RQ2)}

Figure~\ref{fig:exp2} reports the robustness of the four models, trained in the small-network RS setting and tested against unseen adversarial strategies without retraining. Overall, ZT-AIR exhibits the strongest robustness under attacker variation, showing the smallest degradation from the RS baseline across both BFS and DFS.

\begin{icisfigure}{fig:exp2}{Performance (Top) and Action Distributions (Bottom) under Threat Strategies}
\safeincludegraphics[width=6in]{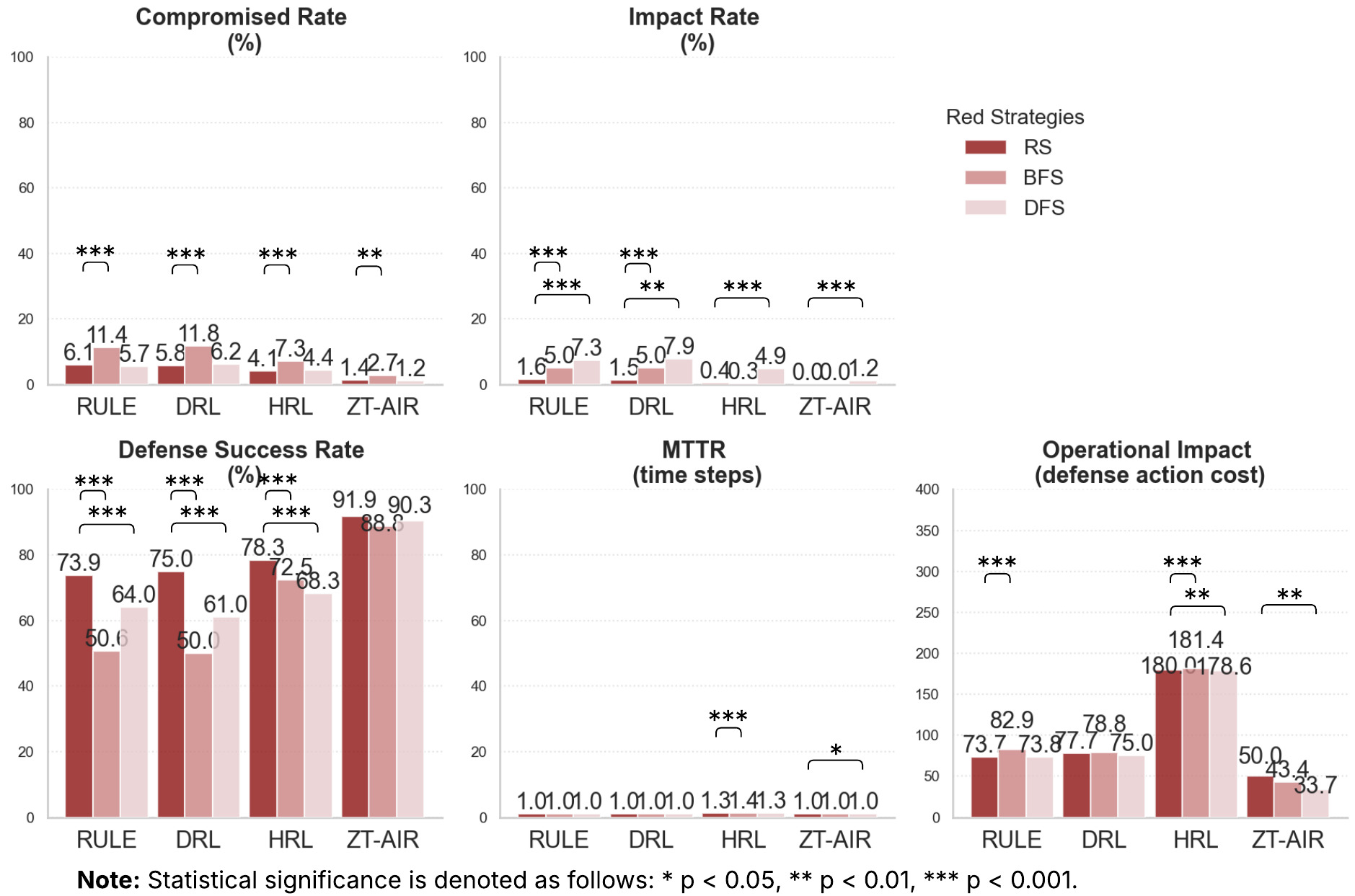}\\[4pt]
\safeincludegraphics[width=3.2in]{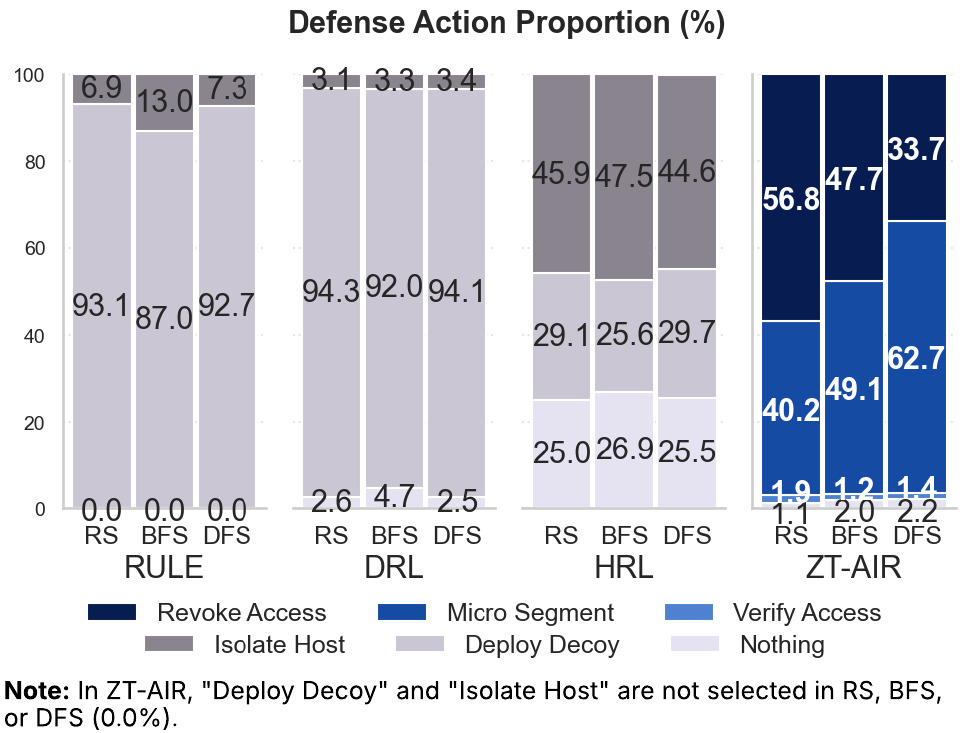}
\end{icisfigure}

\textbf{Baselines: rigidity under distribution shift.} Rule-based, DRL, and HRL all degrade significantly on at least one security metric under both unseen strategies. HRL is more robust in overall security effectiveness, but still degrades substantially when the attacker's traversal logic changes. The action distributions explain these failures, as none of these baselines reallocates its action mix across threat strategies. Rule-based and DRL remain locked on opportunistic deception, whereas HRL remains tied to coarse, costly isolation.

\textbf{ZT-AIR: robustness through action reweighting.} ZT-AIR is not immune to out-of-distribution variation, as its compromised rate rises modestly under BFS and its impact rate rises under DFS, but its degradation remains the smallest among all models. ZT-AIR is the only model whose action distribution shifts with the adversary without retraining. It relies increasingly on \act{micro\_segment} over \act{revoke\_access} as the adversary moves from RS to BFS to DFS. Robustness under attacker variation thus comes not from stronger intervention alone, but from coupling hierarchy with ZTA enforcement mechanisms that can be reweighted as adversary behavior changes.

\subsection{Experiment Set 3: Enterprise-Scale Applicability (RQ3)}

Figure~\ref{fig:exp3} reports the performance of all four models, independently trained and tested across three network scales. ZT-AIR maintains the lowest compromised and impact rates at every scale. It is also the only model whose operational cost decreases at the largest scale.

\begin{icisfigure}{fig:exp3}{Performance (Top) and Action Distributions (Bottom) across Network Scales}
\safeincludegraphics[width=6in]{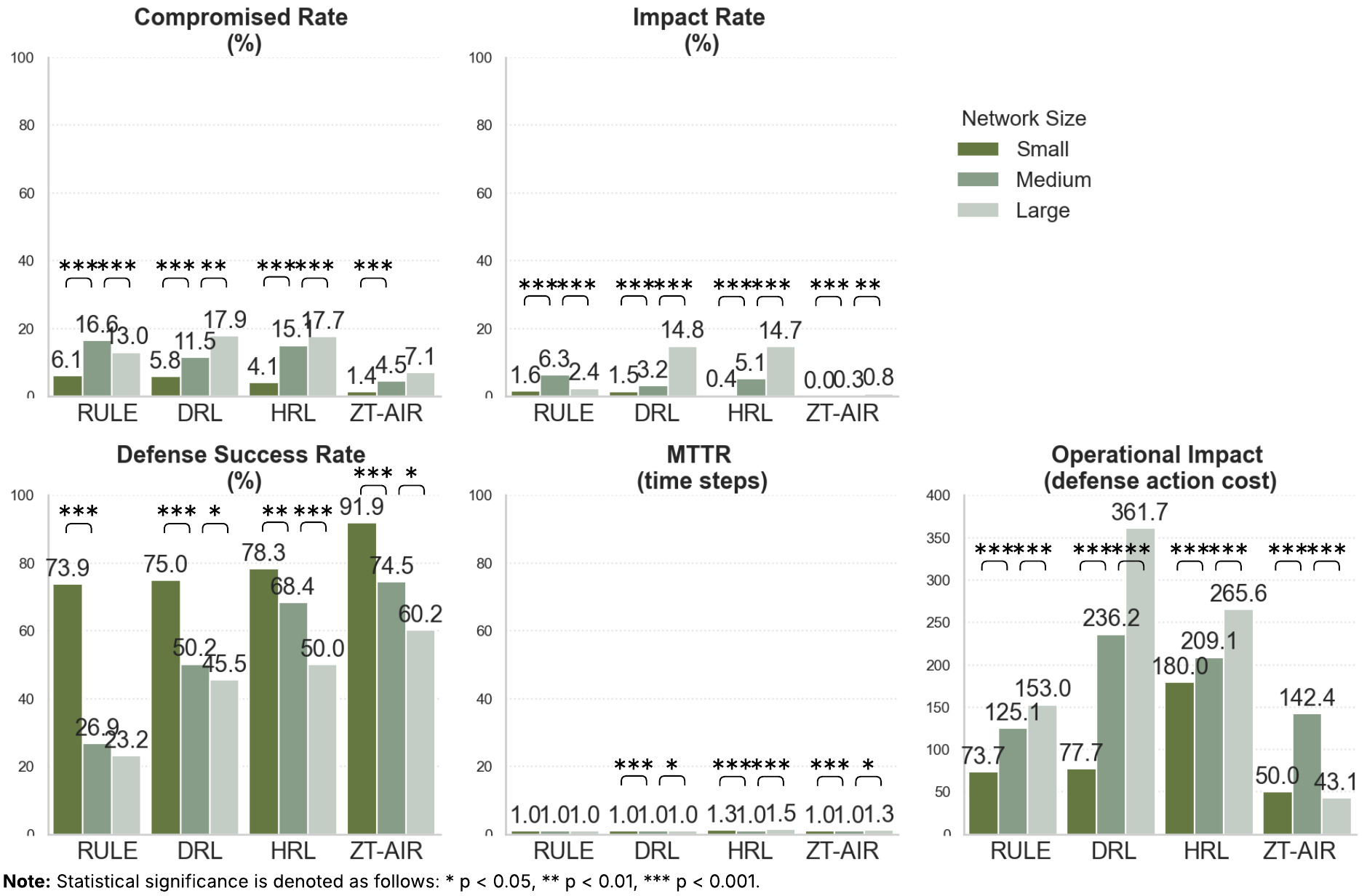}\\[4pt]
\safeincludegraphics[width=3.2in]{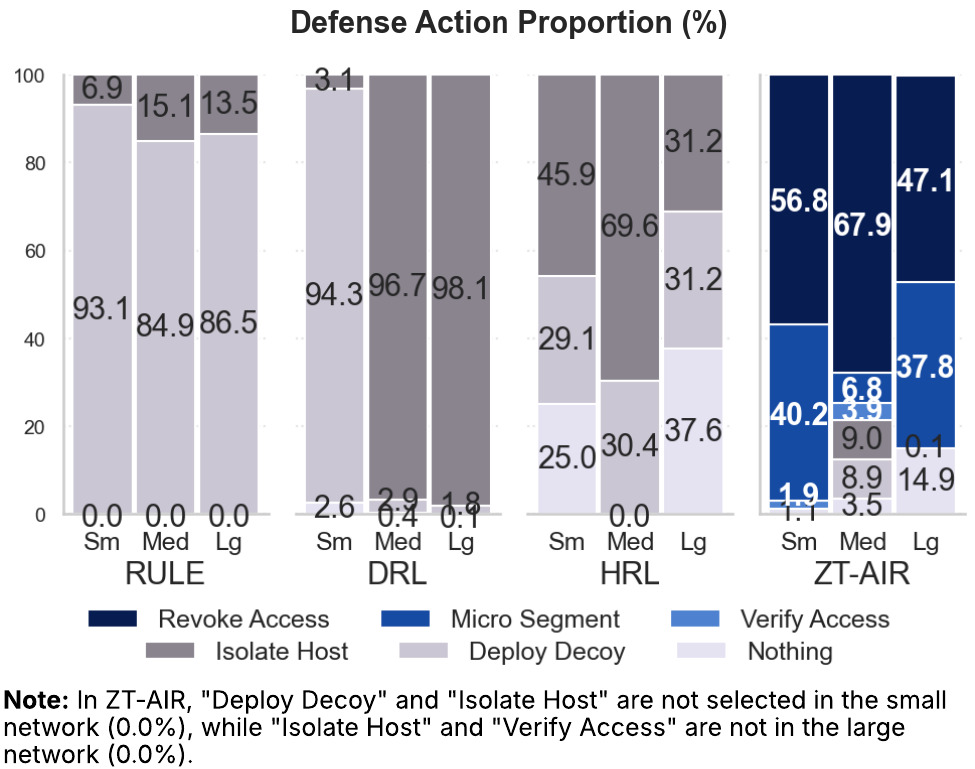}
\end{icisfigure}

\textbf{Baselines: loss of scalable control.} Each baseline fails in a manner tied to its dominant action. Rule-based maintains opportunistic deception, so its effectiveness collapses as decoy coverage becomes diluted across a much larger attack surface. DRL shifts toward isolation, but this reflects indiscriminate coverage by volume rather than precision, so security performance and operational cost worsen together. HRL remains the strongest baseline, but its performance degrades markedly because subnet-level strategic abstraction cannot be translated into accurate host-level targeting as scale grows.

\textbf{ZT-AIR: scalability through action reweighting.} ZT-AIR scales most effectively among all models by maintaining its ZTA action-focused defense logic. Across all scales, ZT-AIR reintroduces conventional actions in limited proportions and shifts toward \act{revoke\_access} over \act{micro\_segment}. ZTA actions remain viable even when exact targeting becomes harder to maintain, because they can be applied preemptively across a broader host set without the costly penalty that makes broad isolation untenable.

\subsection{Experiment Set 4: Operational Generalizability (RQ4)}

Figure~\ref{fig:exp4} compares the performance of the three models, DRL, HRL, and ZT-AIR, independently trained and evaluated in CAGE4. The rule-based approach, which performs poorly in Cyberwheel, is excluded. Overall, ZT-AIR again delivers the strongest aggregate performance on all security metrics, while operational impact, measured as disruption to benign user activity, remains low.

\begin{icisfigure}{fig:exp4}{Performance (Top) and Action Distributions (Bottom) in CAGE4}
\safeincludegraphics[width=6.2in]{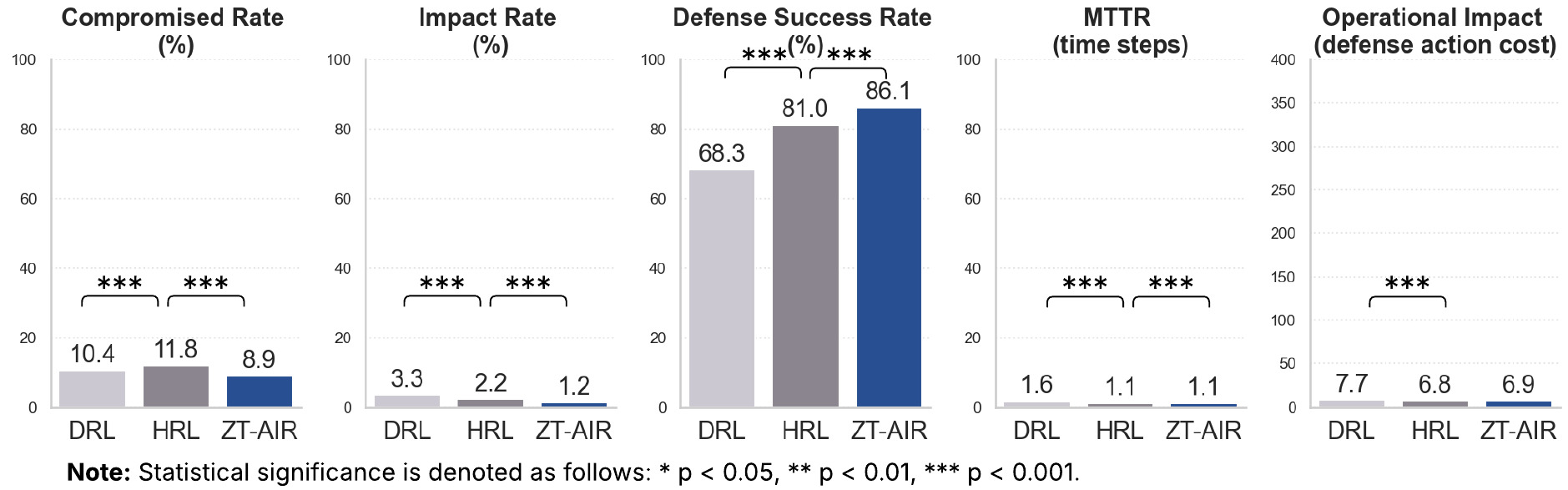}\\[4pt]
\safeincludegraphics[width=3in]{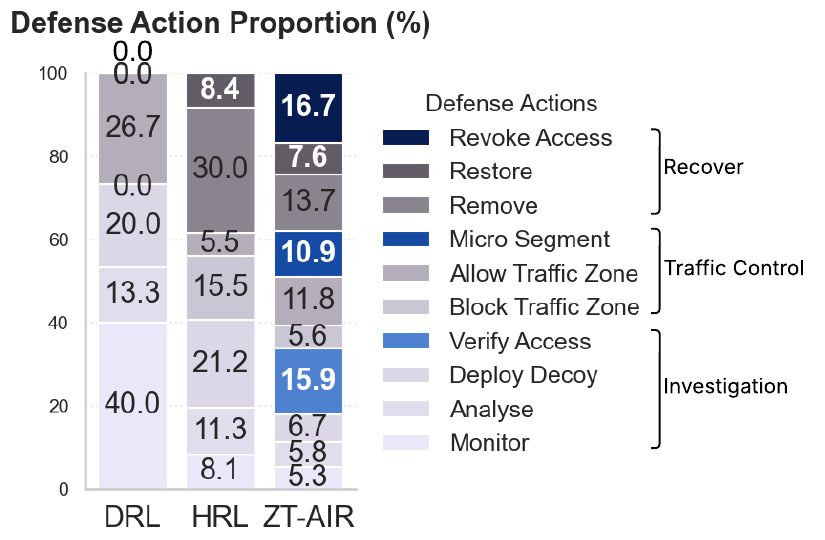}
\end{icisfigure}

\textbf{DRL vs.\ HRL: value of hierarchical decision structure in a distributed defense setting.} The performance gains from hierarchical decision structure carry over from Cyberwheel to CAGE4. HRL allows slightly more initial compromise than DRL, but significantly improves all other security and operational metrics. Figure~\ref{fig:exp4} shows that this improvement originates from resolving the policy collapse of the flat policy. DRL allocates most of its actions to investigation-oriented behaviors and rarely selects direct traffic control or recovery actions. HRL, by contrast, balances its actions across investigation, traffic control, and recovery strategies.

\textbf{HRL vs.\ ZT-AIR: ZTA enforcement mechanisms as fine-grained complements within each strategy.} ZTA enforcement mechanisms again lift the ceiling established by hierarchical decision structure alone. Relative to HRL, ZT-AIR significantly improves all security metrics, while operational impact remains comparably low. This improvement arises from a different composition of actions within each strategy rather than a different strategy distribution. Specifically, \act{deploy\_decoy} is largely displaced by \act{verify\_access} within investigation, subnet-level \act{block\_traffic\_zone} partially by host-level \act{micro\_segment} within traffic control, and \act{remove} partially by immediate credential invalidation via \act{revoke\_access} within recovery. Experiment 4 thus suggests that ZT-AIR is not an artifact tied to a single environment, but a design that generalizes across qualitatively different defense contexts.

\subsection{Operational Plausibility Assessment by SOC Practitioners}

To assess the operational plausibility of our result interpretations, we conducted a structured assessment on the key findings with SOC practitioners through a written survey and a 30-minute debriefing session. They found ZT-AIR's results reasonably plausible with sufficient evidence based on their operational experience. In particular, they strongly endorsed the interpretation that ZT-AIR's joint advantage arises from frequent intervention with low-cost actions rather than stronger blocking. The strongest endorsement concerned the scale experiment. As one practitioner explained, ``broader trust withdrawal---especially for secondary admin accounts---is the most prominent remediation we see in large networks, as it has the least impact on the network and there are natural redundancies in place.'' They also cautioned that the DRL baseline's deception-heavy convergence may depend on the assumed cost of decoy deployment.

\section{Discussion}

\subsection{Practical Implications}

The proposed ZT-AIR artifact translates ZTA from a normative security posture into an executable response capability, creating actionable value for three stakeholder groups.

For security operations practitioners, ZT-AIR addresses a core operational pain point. Existing rule-based playbooks force defenders into a binary choice between cautious monitoring, which allows adversary progression, or aggressive containment, which disrupts legitimate business operations. ZT-AIR reframes this coarse choice by adaptively executing graduated ZTA enforcement actions that intervene early at low cost before stronger escalation.

For chief information security officers (CISOs), ZT-AIR addresses two persistent adoption barriers, namely concerns that stricter trust verification will degrade operational continuity and misperceptions that meaningful ZTA benefits require large-scale deployment and prohibitive investment \parencite{nist2025zta}. Our experiments counter these barriers with empirical evidence of joint gains in security effectiveness and operational continuity that scale across network sizes and generalize to distributed defense settings.

For security technology vendors and researchers, whose products use AI for detection but rely on static playbooks for response, ZT-AIR offers a blueprint for HRL controllers that emit ZTA enforcement actions through environmental interfaces \parencite{nist2020zta}.

\subsection{Contributions to the IS Knowledge Base}

This study addresses a persistent gap in IS DSR on cybersecurity, in that the response execution phase of the IR lifecycle has remained computationally unaddressed. This study, as Type 1 machine learning research \parencite{padmanabhan2022editor}, fills this gap by introducing ZT-AIR as a novel artifact for adaptive IR (a level 1 contribution). It also provides a design-theoretic foundation for how such artifacts can be built (a level 2 contribution) \parencite{gregor2013positioning}.

ZT-AIR's novelty lies in operationalizing ZTA as a design theory for adaptive IR. The artifact translates ZTA's hierarchical structure into a learnable, executable decision logic, encodes enforcement mechanisms as fine-grained, low-disruptive actions, and optimizes security effectiveness together with operational continuity. These integrated design choices yield transferable design principles for adaptive decision-support artifacts operating under uncertainty and competing objectives. Such artifacts should (1) separate strategic prioritization from tactical enforcement through hierarchical decision layers, (2) include fine-grained, lower-disruption controls that complement coarse actions, and (3) optimize primary task effectiveness jointly with operational continuity.

These principles generalize naturally beyond cybersecurity IR. Digital content platforms, for example, face a similar trade-off between safety and legitimate participation. Community- or topic-level prioritization can be separated from post-level intervention, while graduated controls such as reduced amplification, warning labels, and friction-adding review steps provide alternatives to binary takedown. Moderation can thereby optimize harmful-content suppression alongside user engagement.

\section{Conclusion and Future Directions}

Although ZTA has been advanced as the prescribed response to trust erosion, it has remained a normative framework without computational guidance on how to translate its principles into executable defense. This study addressed this gap by developing ZT-AIR, a computational artifact that operationalizes ZTA as a learnable decision logic for adaptive IR through HRL.

Across four experiment sets in Cyberwheel and CAGE4, ZT-AIR consistently achieves the strongest joint performance in security effectiveness and operational continuity. The joint advantage emerges only when hierarchical decision structure is coupled with ZTA enforcement mechanisms, enabling more frequent and less disruptive intervention. It persists under unseen adversarial strategies, increasing network scale, and distributed multi-agent defense. Together, these findings demonstrate that ZTA can be operationalized as a learnable and economically sustainable IR logic rather than a high-level security philosophy.

Future work can extend ZT-AIR's evaluation environment with richer telemetry and, as practitioners suggested, with device criticality, service dependencies, user roles, and business process sensitivity in the cost model. Adversary models can be extended through threat-intelligence-informed attack scenarios and adversarial reinforcement learning. 

\section*{Acknowledgement}
This work was supported by the U.S. Military Academy (USMA) under Cooperative Agreement No. W911NF-25-2-0008. The views and conclusions expressed in this paper are those of the authors and do not reflect the official policy or position of USMA, U.S. Army, U.S. Department of War, or U.S. Government.

\IfFileExists{icis2026_refs.bib}{%
  \bibliographystyle{plainnat}
  \bibliography{icis2026_refs}
}{%
  \typeout{WARNING: references.bib not found; bibliography is not included.}%
}
\end{document}